\documentclass[reprint,superscriptaddress,amsmath,amsfonts,amssymb,aps,prl,showkeys]{revtex4-2}
\usepackage{amsmath}
\usepackage{hyperref}
\usepackage{graphicx}
\hypersetup{
  breaklinks = {true},
  citecolor = {blue},
  colorlinks = {true},
  linkcolor = {red},
}
\usepackage{braket} 

\begin{document}

\title{Proximity Effects Between the Graphene Quasicrystal \\ and Magic-Angle Twisted Bilayer Graphene}

\author{Pedro Alcazar Guerrero}
\affiliation{Catalan Institute of Nanoscience and Nanotechnology (ICN2), CSIC and BIST, Campus UAB, Bellaterra, 08193 Barcelona, Spain}
\affiliation{Department of Physics, Campus UAB, Bellaterra, 08193 Barcelona, Spain}
\author{Viet-Hung Nguyen}
\affiliation{Institute of Condensed Matter and Nanosciences, Universit\'e catholique de Louvain (UCLouvain), 1348 Louvain-la-Neuve, Belgium}
\author{Aron W. Cummings}
\affiliation{Catalan Institute of Nanoscience and Nanotechnology (ICN2), CSIC and BIST, Campus UAB, Bellaterra, 08193 Barcelona, Spain}
\author{Jean-Christophe Charlier}
\affiliation{Institute of Condensed Matter and Nanosciences, Universit\'e catholique de Louvain (UCLouvain), 1348 Louvain-la-Neuve, Belgium}
\author{Stephan Roche}
\affiliation{Catalan Institute of Nanoscience and Nanotechnology (ICN2), CSIC and BIST, Campus UAB, Bellaterra, 08193 Barcelona, Spain}
\affiliation{ICREA--Instituci\'o Catalana de Recerca i Estudis Avan\c{c}ats, 08010 Barcelona, Spain}

\begin{abstract}
We present a numerical study of three-layer graphene heterostructures in which the layers are twisted by the magic angle ($\sim$1.1$^\circ$) or by $\sim$$30^\circ$ to form a graphene quasicrystal. The heterostacks are described using realistic structural relaxations and tight-binding Hamiltonians, and their transport properties are computed for both pristine and disordered systems containing up to $\sim$8 million atoms. Owing to the weak interlayer coupling, we resolve the hybridization between magic-angle flat bands and quasicrystalline states, which are modified in distinct ways across low- and high-energy windows, revealing a new hybrid electronic regime to explore.
\end{abstract}

\maketitle

\textit{Introduction}. The fascinating discovery of the superconducting and Mott insulating phases in magic-angle twisted bilayer graphene (MATBLG) has generated a huge interest in deepening our understanding of the physics of flat-band systems, where a dominant Coulomb interaction triggers a wealth of emerging many-body phenomena \cite{Cao2018, Cao2018b, Stepanov2020, pnas1108174108, Andrei2021, Tilak2021, Serlin2020, Saito2020, Balents2020, PhysRevLett.127.197701, Xie2021, Pierce2021, PhysRevLett.126.137601, Park2022, Jaoui2022, Paul2022, Klein2023}. In parallel, a ``graphene quasicrystal" for a twist angle of 30$^{\circ}$ between two graphene layers has been observed \cite{Ahn2018}, and later grown deterministically via chemical vapor deposition \cite{Pezzini2020}, eliminating the need for artificial assembly. In the decades after their first experimental observation \cite{PhysRevLett.53.1951}, the realization of quasicrystalline structures has captivated material scientists for their quasiperiodic tiling with self-similar properties \cite{PhysRevLett.53.2477,Roche1997,Macia2006, Bindi2009, Bindi2023}. Very recently, other generalizations of Moiré quasicrystals have been also discovered \cite{Uri2023, lai2023imaging}.

MATBLG is dominated by low-energy flat bands, whereas graphene quasicrystals display their most distinctive states at high energies, with low-energy behavior resembling that of isolated graphene layers. Their combination could create a nontrivial hybrid of these two regimes, both still actively explored.

The graphene quasicrystal is by definition aperiodic, which complicates the study of its electronic properties. The juxtaposition of MATBLG with a graphene quasicrystal presents further computational issues, owing to the absence of translational invariance combined with the large unit cell of MATBLG. One can construct quasicrystalline approximant structures that are periodic but which share electronic similarities with the aperiodic quasicrystal \cite{Moon2019, Yu2019}, but the system size quickly increases to millions of atoms as the chosen angle approaches 30$^\circ$. This makes a theoretical exploration of those structures, as well as their disordered variants, out of the reach of conventional computational approaches, hence limiting the inspection of possible undiscovered properties of ``MATBLG/graphene quasicrystal" hybrid stacks.

Here we overcome these difficulties by implementing a realistic tight-binding model which describes the electronic properties of graphene layers stacked at both the magic angle and the quasicrystalline approximant twist angles. In combination with an efficient linear-scaling approach \cite{FAN20211}, we simulate the electronic transport properties of {the graphene quasicrystal and the} MATBLG/graphene quasicrystal trilayer stacks containing up to $\sim$8 million atoms, thus allowing for the first time the study of the respective modifications of flat bands and quasicrystalline-like states driven by proximity effects.

In the absence of disorder, the flat bands of MATBLG are found to be {unaffected} by the quasicrystalline states. This contrasts with the high-energy quasicrystalline states, which are severely suppressed by the interaction with MATBLG. However, {this changes} when introducing disorder. In this case, in the low-energy flat bands we observe a reversal of the recently discovered disorder-induced delocalization in MATBLG \cite{guerrero2024disorderinduced}. Meanwhile, in the high-energy quasicrystalline states,  {anomalous (sub-ballistic) transport properties (typical to pristine quasiperiodic ordering) are always suppressed by disorder, independent of} the presence of the magic-angle graphene layer.
{Overall, our results indicate that a third graphene layer is generally detrimental to observing the anomalous transport features of both the low-energy flat bands in MATBLG and the high-energy quasicrystal states in disorder-free $30^\circ$-BLG.}

\begin{figure}[t]
\includegraphics[width=\linewidth]{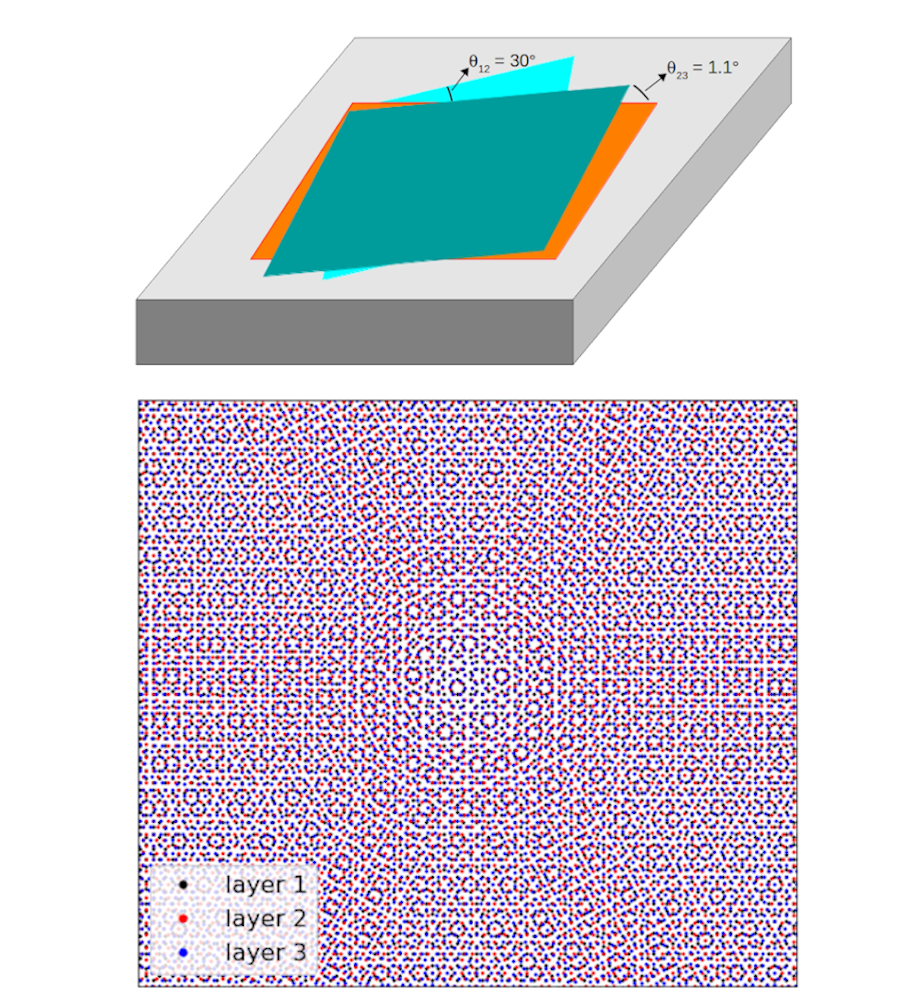}
\caption{Schematic of twisted multilayer graphene systems on a flat substrate such as bulk h-BN (top panel), and top view of the atomic structure of the trilayer stack (bottom panel). Patterns reminiscent of the AA (center region) and AB zones of MATBLG and the 12-fold rotational structures of the quasicrystalline approximant are visible.}
\label{fig1}
\end{figure}

\textit{Structural optimization and tight-binding model}. The atomic structures of graphene superlattices at the magic angle and at the quasicrystal approximant angles were relaxed using molecular dynamics simulations with classical potentials \cite{Hung2021, Nguyen_2022}. Intralayer forces were computed using the optimized Tersoff potentials \cite{Lindsay2010}, whereas interlayer forces were modeled using the Kolmogorov–Crespi potentials \cite{PhysRevB.71.235415, Leven2016}. The atomic structures were optimized until all force components were smaller than 0.5 meV/\AA. To model quasicrystalline graphene, we consider two approximant angles close to 30$^\circ$, namely 31$^\circ$ and 29.8$^\circ$ \cite{Moon2019}. These approximants share similarities with the fully aperiodic quasicrystal, such as the formation of high-energy states whose real space projections exhibit dodecagonal symmetry \cite{Ahn2018}.

Twisted graphene systems are often deposited on a substrate which could have a flat surface, such as h-BN \cite{Dean2010} (see Fig.\ \ref{fig1}). The interaction between the flat substrate and the bottom graphene layer can diminish the moiré-induced corrugation in this layer. For this reason, both free-standing graphene and graphene on a substrate are investigated. In the former case, all C-atoms can freely move during the relaxation calculation. In the latter case, C-atoms can move under the condition that the bottom graphene layer is kept flat.
In our simulations of electronic properties, we have found a negligible difference between these structures.

The electronic properties of twisted graphene systems are then computed using the $p_z$ tight-binding (TB) Hamiltonian elaborated in \cite{Hung2021, Nguyen_2022}. The hopping energies $t_{nm}$ between carbon $n$- and $m$-sites are determined by the standard Slater–Koster formula
$$ t_{nm} = \cos^2 \phi_{nm} V_{pp\sigma} (r_{nm}) + \sin^2 \phi_{nm} V_{pp\pi} (r_{nm}),$$
where the direction cosine of $\vec r_{nm} = \vec r_{m} - \vec r_{n}$ along the $z$ axis is $\cos\phi_{nm} = z_{nm}/r_{nm}$. The distance-dependent Slater-Koster parameters are determined as \cite{Trambly2010, Trambly2012} 
\begin{equation*}
\begin{aligned}
& V_{pp\pi} (r_{nm}) = V_{pp\pi}^0 \exp \left[ q_\pi \left( 1 - \frac{r_{nm}}{a_0}  \right) \right] F_c (r_{nm}),  \\
& V_{pp\sigma} (r_{nm}) = V_{pp\sigma}^0 \exp \left[ q_\sigma \left( 1 - \frac{r_{nm}}{d_0}  \right) \right] F_c (r_{nm}),
\end{aligned}
\end{equation*}
with a smooth cutoff function $F_c (r_{nm}) = \left[ 1 + \exp \left(( r_{nm}-r_c)/\lambda_c\right) \right]^{-1}$. Note that here we use a cutoff of electronic couplings, $r_\mathrm{inplane} \leq 0.432$ nm \cite{Nguyen_2022}. This approximation results in a highly sparse TB Hamiltonian matrix, making calculations feasible for small twist angles and/or multilayered systems with a large number of atoms. Within this approximation, the adjusted TB parameters are $V_{pp\pi}^0 = -2.7$ eV, $V_{pp\sigma}^0 = 367.5$ meV, $\frac{q_\pi}{a_0} = \frac{q_\sigma}{d_0} = 22.18$ nm$^{-1}$, $a_0 = 0.1439$ nm, $d_0 = 0.33$ nm, $r_c = 0.614$ nm, and $\lambda_c = 0.0265$ nm. To model disorder, a random Anderson potential uniformly distributed within $[-W/2,+W/2]$ is added to the onsite energies of the $p_z$ orbitals.

\begin{figure}[t]
\includegraphics[width=\linewidth]{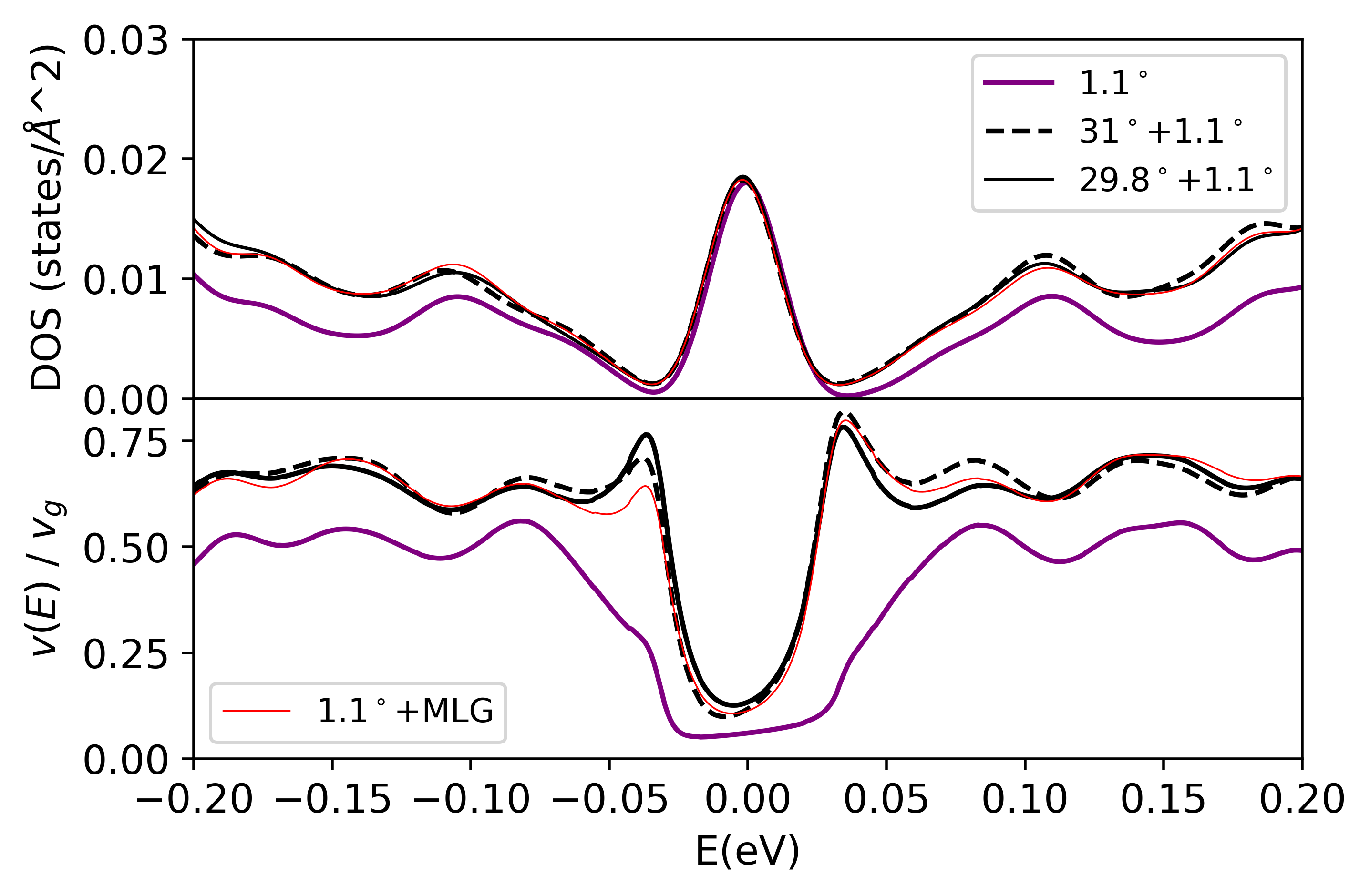}
\caption{Density of states (top panel) and Fermi velocities (bottom panel) of MATBLG (purple solid line) and the two trilayer systems (black solid and dashed lines). {The averaged Fermi velocity of the trilayer system is depicted by the red line, considering a trilayer system where the MATBLG and the quasicrystalline layer are decoupled (see main text).}}
\label{fig2}
\end{figure}

\textit{Calculation of electronic properties}. We use an efficient linear-scaling quantum transport methodology \cite{FAN20211} to study the electronic properties of the MATBLG/quasicrystal trilayer stacks {and the two involved bilayers}. For transport, the main quantity we calculate is the mean square displacement of an initial state $\ket{\psi(0)}$, given by
\begin{equation*}
\Delta X^2(E,t) = \frac{ \langle\psi_X(t) | \delta(E-\mathcal{\hat H}) | \psi_X(t)\rangle }{ \rho(E) },
\end{equation*}
where $\mathcal{\hat H}$ is the TB Hamiltonian, $\ket{\psi_X(t)} = [ \hat{X},\hat{U}(t) ] \ket{\psi(0)}$, $\hat{X}$ is the position operator, $\hat{U}(t) = \exp(-\mathrm{i}\mathcal{\hat  H}t/\hbar)$ is the time evolution operator, and $\rho(E) = \braket{\psi(0) | \delta(E-\mathcal{\hat H}) | \psi(0)}$ is the density of states (DOS). To avoid direct diagonalization, the time evolution operator and the energy projection operator $\delta(E-\mathcal{\hat  H})$ are expanded as a series of Chebyshev polynomials. The initial state $\ket{\psi(0)}$ is a random-phase state that permits efficient calculation of electronic properties over the full Hamiltonian spectrum \cite{WeisseKPM}. From the mean square displacement we extract the diffusion coefficient $D(E,t)= \frac{1}{2}\frac{\partial}{\partial t}\Delta X^2(E,t)$. In the ballistic regime we extract the Fermi velocity via $D(E,t) = v^{2}(E)t$, while with disorder the mean free path can be deduced from $\ell(E) = 2D_\mathrm{max}(E)/v(E)$, where $D_\mathrm{max}(E) = \max\limits_t D(E,t)$.

We use timesteps of 10 and 1 fs for transport in the clean and disorderd systems, respectively. We expand $\delta(E-\mathcal{\hat  H})$ as a series of 3500 Chebyshev polynomials with the Jackson kernel \cite{WeisseKPM}, corresponding to a Gaussian energy broadening of $\sim$23 meV. The trilayer systems contain a total of $\sim$5 million atoms for the $31^\circ$ approximant and $\sim$8 million atoms for the $29.8^\circ$ approximant.

\textit{Low-energy regime and flat-band states}. We first examine the low-energy flat-band states in MATBLG, and how they are modified by adding a $\sim$$30^\circ$-rotated graphene layer.
In Fig.\ \ref{fig2} (top panel), we plot the DOS of MATBLG and each of the two MATBLG/quasicrystal approximants, focusing on the low-energy regime. In the flat-band region, $|E| \lesssim 25$ meV, the DOS is nearly identical for all three cases, while a small increase is seen at larger energies for the trilayer structures.

In the bottom panel of Fig.\ \ref{fig2}, we plot the Fermi velocity of each system. Here we see a substantial modification of ballistic transport between the bi- and trilayer cases, manifesting as a significant increase of the Fermi velocity (up to $5\times$) in the trilayer case. However, our methodology calculates the Fermi velocity averaged over the entire Fermi surface, and this result may thus arise from the average of parallel transport in the MATBLG layers at velocity $v_\mathrm{MA}$ and a decoupled quasicrystalline layer at $v_\mathrm{g}$. {To check this, we have made calculations for a decoupled system by removing all interlayer interactions between the MATBLG and the bottom graphene layer. This corresponds to the red curve in Fig.\ \ref{fig2}, where we see no significant difference arise at low energy for the DOS nor for the Fermi velocity.}

\begin{figure}[t]
\includegraphics[width=\linewidth]{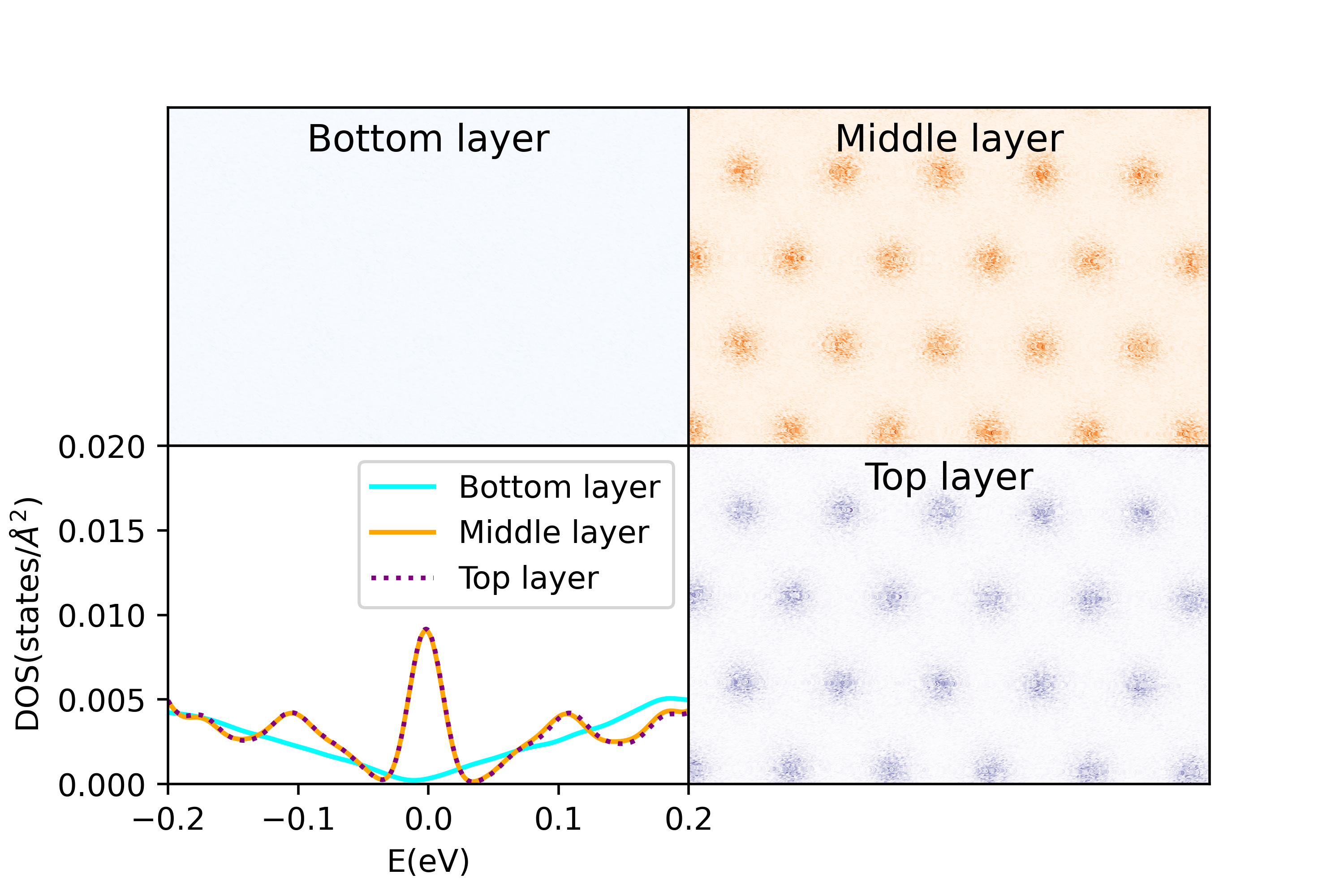}
\caption{Layer-projected DOS (bottom left panel) and LDOS of each of the layers for the $31^\circ$+$1.1^\circ$ trilayer hybrid system at charge neutrality ($E=0$). The moiré pattern is visible in the two $1.1^\circ$-rotated layers, while the extra quasicrystalline layer does not display any signal of the flat band.}
\label{fig3}
\end{figure}

This observation is consistent with the layer-projected DOS and local DOS (LDOS) in each of the layers, shown in Fig.\ \ref{fig3}. Here we see that the DOS of the MATBLG layers (middle and top layers) exhibits the flat-band peak at $E = 0$, while the quasicrystal layer (bottom layer) resembles single-layer graphene. Similarly, the LDOS of the MATBLG layers exhibits the characteristic moiré periodicity, with states localized in the AA overlap regions, while the quasicrystal layer shows no evidence of the moiré superlattice.

\begin{figure}[t]
\includegraphics[width=\linewidth]{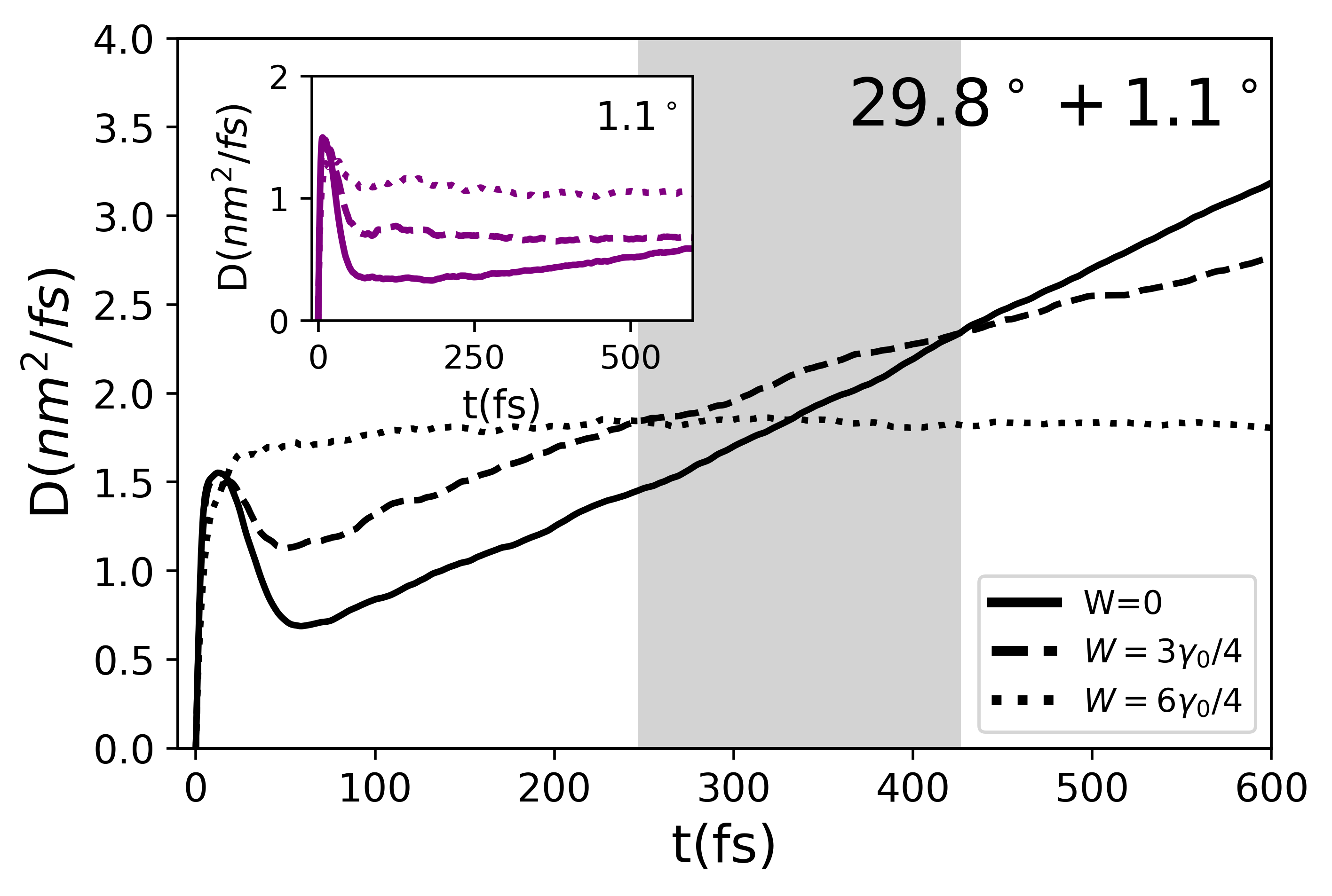}
\caption{Time-dependent diffusion coefficient in the flat band for different disorder strengths, for MATBLG (inset) and the trilayer with the 29.8º approximant (main panel).
}
\label{fig5}
\end{figure}

The above results have shown that, in the clean limit, adding a $\sim$$30^\circ$-rotated graphene layer to MATBLG has little impact on the low-energy flat-band states. We now see if this extends to transport in the presence of disorder. We consider Anderson disorder values, $W = 3V^0_{pp\pi}/4$ and $6V^0_{pp\pi}/4$, where we see an anomalous disorder-induced delocalization in MATBLG, as discovered in Ref.\ \citenum{guerrero2024disorderinduced}. We reproduce these results in the inset of Fig.\ \ref{fig5}, where we show the time evolution of the diffusion coefficient of MATBLG in the middle of the flat band ($E=0$). Here we see that increasing the disorder strength leads to an anomalous increase in $D$ at long times.

This situation is strongly altered in the trilayer case, as shown in the main panel of Fig.\ \ref{fig5}. Here we see that at long times, increasing the disorder strength always reduces the diffusion coefficient. The anomalous disorder-induced delocalization seen in MATBLG is thus suppressed when a quasicrystalline graphene layer is added to the stack, resulting in ``standard'' transport behavior. This occurs despite the strong resilience of the state projection on the MATBLG layers in the clean case. {The presence of disorder thus appears to enhance the coupling between the layers and wash out the flat bands by enabling scattering between the flat-band states and higher-energy states in the third layer.}

\begin{figure}[t]
\includegraphics[width=\linewidth]{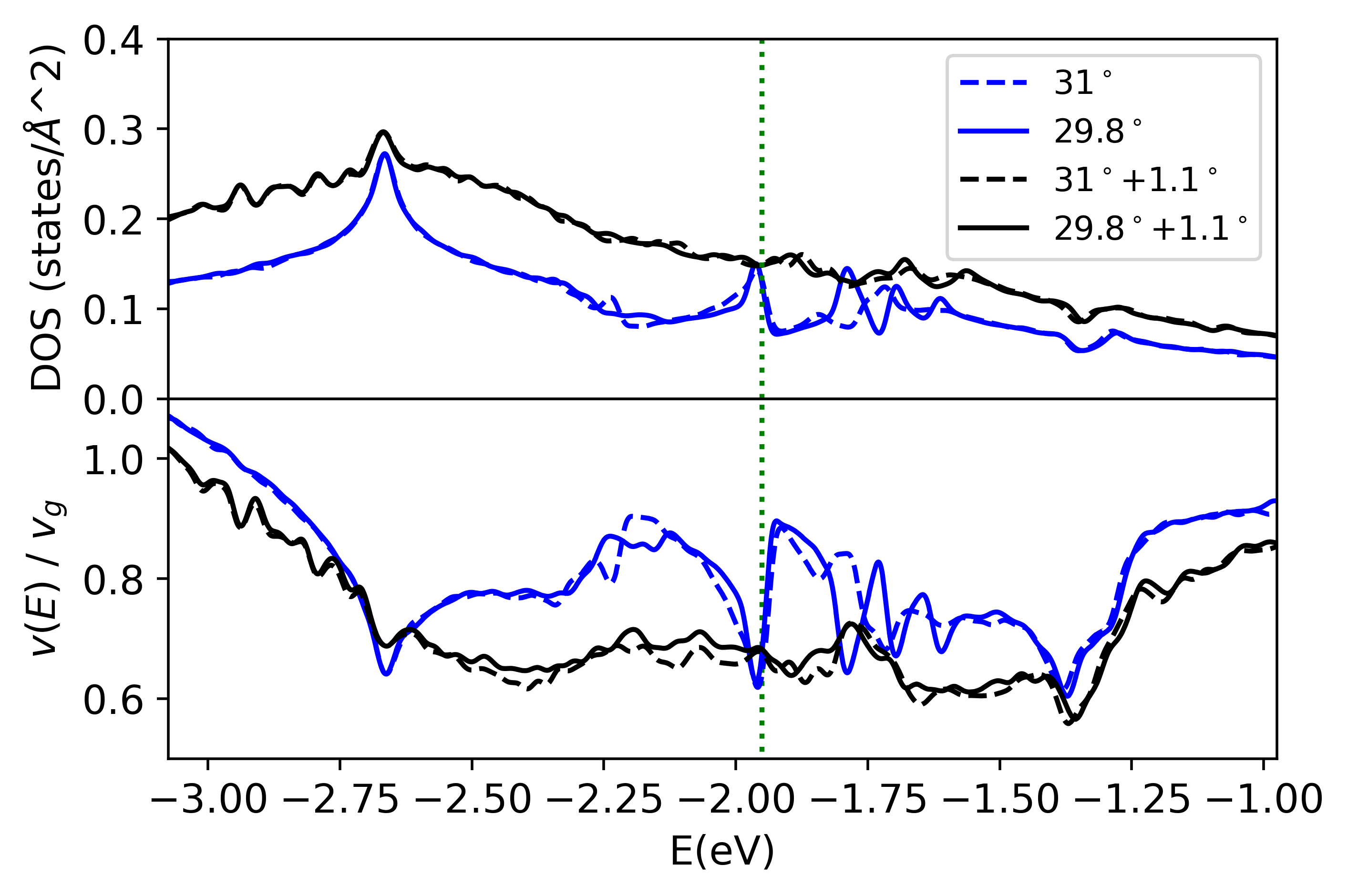}
\caption{DOS (top panel) and Fermi velocity (bottom panel) in the high-energy quasicrystalline-state regime. The vertical green dotted line marks the energy of the $\alpha$ quasicrystalline peak ($E=-1.95$ eV) \cite{Moon2019}.}
\label{fig4}
\end{figure}

\textit{High-energy regime and quasicrystalline states}. We now focus on the electronic structure and transport in the high-energy regime. Here we obtain a favorable comparison to prior results on graphene quasicrystalline approximants \cite{Crosse2021, Moon2019, 10.3389/frcrb.2024.1496179}, which have demonstrated the presence of resonant peaks in the DOS at energies between $-2$ and $-1.5$ eV, including the so-called $\alpha$-peak shown as the vertical dotted line in Fig.\ \ref{fig4}. These peaks correspond to localized states lacking spatial periodicity and displaying quasiperiodic orientational patterns. The top panel of Fig.\ \ref{fig4} shows that adding an extra $1.1^\circ$-rotated layer on top of the quasicrystalline stack leads to strong suppression of the quasicrystalline fingerprints in the DOS. 

Figure \ref{fig4} (bottom panel) shows the corresponding normalized Fermi velocity of each structure. Consistent with the suppression of the resonant peaks in the DOS, the velocity profile also becomes relatively featureless with the addition of the $1.1^\circ$-rotated layer. Overall, the additional layer induces a reduction of the Fermi velocity throughout the entire energy range of interest. 

\begin{figure}[t]
    \centering
    \includegraphics[width=\linewidth]{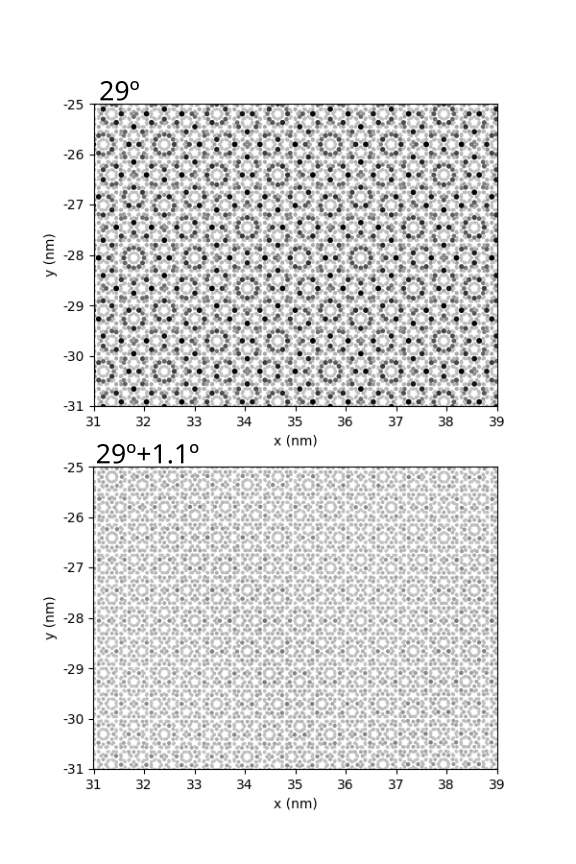}
    \caption{{Local density of states of $29.8^\circ$-twisted bilayer graphene (top) and the two bottom layers of the $29.8^\circ + 1.1^\circ$ trilayer at the $\alpha$ peak (vertical dotted line in Fig.\ \ref{fig4}). The fractal localization pattern is destroyed by the addition of the extra layer and replaced with a homogeneously distributed local density of states.} }
    \label{fig:LDOS}
\end{figure}

{The LDOS of the pristine quasicrystal bilayer, shown in Fig.\ \ref{fig:LDOS} (top), indicates the emergence of the fractal localization pattern, consistent with prior results \cite{Ahn2018,Moon2019, 10.3389/frcrb.2024.1496179}. In contrast, the LDOS of the equivalent atomic coordinates in the trilayer does not exhibit any anomalous localization features, indicating that the addition of an extra layer destroys the quasicrystalline states and their related features.}

\begin{figure}[t]
\includegraphics[width=\linewidth]{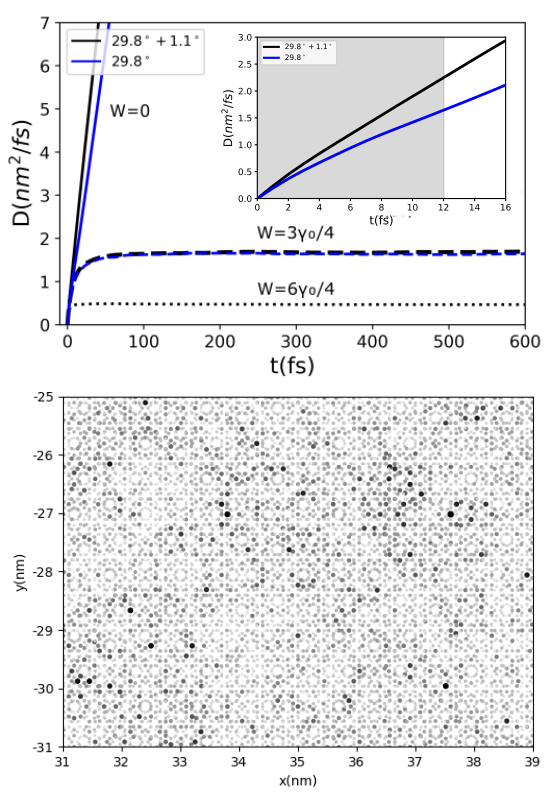}
\caption{{Top panel: time-dependent diffusion coefficient of the quasicrystalline approximant and the trilayer system at the energy of the $\alpha$-peak (vertical dotted line in Fig.\ \ref{fig4}). Inset: zoom of the short-time diffusion coefficient, indicating the sub-ballistic behavior of the quasicrystal (blue line). The shadowed region corresponds to times where the mean squared displacement has not reached the periodicity of the approximant. Bottom panel: local density of states of the $29.8^\circ$ approximant with Anderson disorder strength $W=3\gamma_0/4$, indicating suppression of the fractal localization pattern.}}
\label{fig6}
\end{figure}

Finally, we focus on transport in the high-energy regime, where the formation of resonant quasicrystalline states has been observed experimentally and is well reproduced theoretically. At those resonance energies, the real-space extension of the electronic states indicates the emergence of dodecagonal ordering (incompatible with translational invariance), which could point towards unconventional electronic features in the limit of a totally aperiodic structure, as known for other quasicrystalline structures \cite{Roche1997, PhysRevB.104.144201, PhysRevB.104.155304, peng2024structural, PhysRevB.107.054206, PhysRevLett.132.086402,HectorPRL1}. {We can spot signals of this unconventional transport in the short time behavior shown in the inset of Fig.\ \ref{fig6} (top panel). Here, the quantum dynamics in the quasicrystalline approximant (indicated by the blue line) exhibit the typical sub-ballistic transport even in the absence of disorder, as observed in pure quasiperiodic systems \cite{Roche1997, RochePRL97, Roche98, PhysRevB.69.121410, PhysRevB.65.220202, Trambly_de_Laissardi_re_2026}.}
{However, adding the third $1.1^\circ$-layer restores ballistic transport in the absence of disorder (black line), consistent with the suppression of quasiperiodic-induced anomalous localization effects indicated in Figs.\ \ref{fig4} and \ref{fig:LDOS}}.

Additionally, as shown in the top panel {of Fig.\ \ref{fig6}}, Anderson disorder leads to standard diffusive behavior {for both the bilayer quasicrystal and the trilayer}, with a stronger disorder leading to slower diffusion and a reduced mean free path. {The bottom panel shows the LDOS of the $29.8^\circ$ approximant with a superimposed Anderson disorder of strength $W=3V_{pp\pi}^0/4$. The fractal localization pattern present in the pristine case (top panel of Fig.\ \ref{fig:LDOS}) is clearly destroyed}. {These results, as well as those of the clean case, suggest that the graphene quasicrystal states are poorly resilient against local perturbation, whether arising from short-range electrostatic disorder or from coupling to another layer.}


\textit{Summary and discussion}. We have shown that interfacing MATBLG with graphene quasicrystal approximants leads to a variety of proximity effects. While the low-energy flat-band states appear to be insensitive to the addition of a $\sim$$30^\circ$-rotated layer in the clean case, the high-energy quasicrystalline states are suppressed by the presence of a magic-angle layer.
{Meanwhile, adding disorder reveals a couple different features. In the trilayer, the presence of disorder always leads to standard diffusive behavior, both in the low-energy flat bands and in the high-energy quasicrystalline states. Disorder also washes out the quasicrystalline states in the bilayer system, while an anomalous increase of mean free path with disorder strength is observed in the flat bands of MATBLG \cite{guerrero2024disorderinduced}.
}

{These results have a couple implications for heterostructures involving flat-band or quasicrystal states in twisted BLG. In the case of MATBLG, Figs.\ \ref{fig2} and \ref{fig3} suggest that a $30^\circ$-rotated graphene layer can serve as a protective capping layer to protect flat-band states from extrinsic disorder or adsorbates. However, this holds only in the clean case. Once disorder is introduced, the anomalous features of the flat-band states are washed out, indicating that such a capping layer is ultimately more detrimental than beneficial. A semiconducting or insulating capping layer may be more useful for this purpose, as a large band gap may prevent disorder-induced mixing of the flat bands with higher-energy states of the capping layer.}

{In the quasicrystalline case, we find that the graphene quasicrystal state is destroyed both by disorder and by interaction with a third graphene layer, suggesting that it is highly fragile to local perturbations. This contrasts with the physics of intermetallic alloy-based quasicrystals, where disorder-induced delocalization remains a long-standing conundrum \cite{RochePRL97, Roche1997, Roche98}. Here a difference may lie in the short- vs. long-range nature of the disorder; as the quasicrystal state exhibits localized features on the scale of the graphene lattice constant \cite{Moon2019}, it may be more susceptible to the short-range disorder we consider here. Meanwhile, long-range disorder, such as that arising from electron-hole puddles \cite{Xue2011puddles}, may preserve the anomalous transport features of the quasicrystalline state.}

{A similar, but opposite, argument may be made in the case of MATBLG. Here the flat band states have a periodicity of $14$ nm \cite{guerrero2024disorderinduced}, and short-range disorder may therefore average out over this length scale and be less effective. Meanwhile, electron-hole puddles, whose length scale is similar to the periodicity of the magic-angle Moiré \cite{Xue2011puddles}, may be more effective in suppressing the anomalous transport features of the flat bands.}

{To summarize, our simulations indicate that a third graphene layer is generally detrimental to observing the anomalous transport features of both the low-energy flat bands in MATBLG and the high-energy quasicrystal states in $30^\circ$-BLG. Looking ahead, our results open promising avenues for exploring these systems, particularly their response to the diverse types of defects and disorder commonly encountered in bilayer graphene, as well as the extension of proximity effects to other types of magnetic or strong spin-orbit coupling materials.}



\begin{acknowledgments}
ICN2 is funded by the CERCA Programme/Generalitat de Catalunya and supported by the Severo Ochoa Centres of Excellence programme, Grant CEX2021-001214-S, funded by MCIN/AEI/10.13039.501100011033.  This work is also supported by MICIN with European funds‐NextGenerationEU (PRTR‐C17.I1) and by 2021 SGR 00997, funded by Generalitat de Catalunya.
V.-H.N. and J.-C.C. acknowledge financial support from the F\'ed\'eration Wallonie-Bruxelles through the ARC Grant ``DREAMS'' (No.~21/26-116), from the EOS project ``CONNECT'' (No.~40007563), from the EU Pathfinder project ``FLATS'' (No.~101099139), and from the Belgium F.R.S.-FNRS through the research project (No.~T.029.22F). 
Simulations were performed at the Center for Nanoscale Materials, a U.S.\ Department of Energy Office of Science User Facility, supported by the U.S.\ DOE, Office of Basic Energy Sciences, under Contract No.\ 83336 {and 83864}. Computational resources were also provided by the supercomputing facilities of UCLouvain (CISM) and the Consortium des Équipements de Calcul Intensif (C\'ECI), funded by the F.R.S.-FNRS under Grant No. 2.5020.11 and by the Walloon Region. 
\end{acknowledgments}

\bibliography{references}

\end{document}